\documentclass{emulateapj}

\usepackage{float}
\usepackage{amsmath}
\usepackage{epsfig,floatflt}
\usepackage{subfigure}

\begin{document}

\title{Bayesian analysis of the low-resolution polarized 3-year WMAP
  sky maps}

\author{H.\ K.\ Eriksen\altaffilmark{1,2,3}, Greg
  Huey\altaffilmark{4,5,6}, A. J. Banday\altaffilmark{7}, K. M.
  G\'{o}rski\altaffilmark{5,6,8}, J. B.  Jewell\altaffilmark{4,5}, \\I.\
  J.\ O'Dwyer\altaffilmark{4,5}, B.\ D.\ Wandelt\altaffilmark{4,7,9}}

\altaffiltext{1}{email: h.k.k.eriksen@astro.uio.no}

\altaffiltext{2}{Institute of Theoretical Astrophysics, University of
Oslo, P.O.\ Box 1029 Blindern, N-0315 Oslo, Norway}

\altaffiltext{3}{Centre of
Mathematics for Applications, University of Oslo, P.O.\ Box 1053
Blindern, N-0316 Oslo}

\altaffiltext{4}{Department of Physics, University of Illinois,
  Urbana, IL 61801}

\altaffiltext{5}{Jet Propulsion Laboratory, 4800 Oak
  Grove Drive, Pasadena CA 91109} 

\altaffiltext{6}{California Institute of Technology, Pasadena, CA
  91125} 

\altaffiltext{7}{Max-Planck-Institut f\"ur Astrophysik,
Karl-Schwarzschild-Str.\ 1, Postfach 1317, D-85741 Garching bei
M\"unchen, Germany}

\altaffiltext{8}{Warsaw University Observatory, Aleje Ujazdowskie 4, 00-478 Warszawa,
  Poland}

\altaffiltext{9}{Astronomy Department, University of Illinois at
  Urbana-Champaign, IL 61801-3080}

\date{Received - / Accepted -}

\begin{abstract}
We apply a previously developed Gibbs sampling framework to the
foreground corrected 3-yr WMAP polarization data and compute the power
spectrum and residual foreground template amplitude posterior
distributions. We first analyze the co-added Q- and V-band data, and
compare our results to the likelihood code published by the WMAP
team. We find good agreement, and thus verify the numerics and data
processing steps of both approaches. However, we also analyze the Q-
and V-band separately, allowing for non-zero EB cross-correlations and
including two individual foreground template amplitudes tracing
synchrotron and dust emission. In these analyses, we find tentative
evidence of systematics: The foreground tracers correlate with each of
the Q- and V-band sky maps individually, although not with the
co-added QV map; there is a noticeable negative EB cross-correlation
at $\ell \lesssim 16$ in the V-band map; and finally, when relaxing
the constraints on EB and BB, noticeable differences are observed
between the marginalized band powers in the Q- and V-bands. Further
studies of these features are imperative, given the importance of the
low-$\ell$ EE spectrum on the optical depth of reionization $\tau$ and
the spectral index of scalar perturbations $n_{s}$.
\end{abstract}


\maketitle

\section{Introduction}

One of the most remarkable results in the 3-yr data release from the
Wilkinson Microwave Anisotropy Probe (WMAP) experiment
\citep{hinshaw:2007, page:2007} was the detection of large-scale
E-mode polarization at millimeter wavelengths. This was interpreted as
the theoretically predicted signature of reionization, and allowed the
WMAP team to set new and tighter constraints on the optical depth of
reionization $\tau$. In turn, the well-known degeneracy between $\tau$
and the spectral index of primordial scalar perturbations $n_{s}$ was
broken. The final outcome was a claimed detection of $n_{s} \ne 1$ at
a statistical significance of almost $3\sigma$ \citep{spergel:2007}.

One should bear in mind, however, the great potential for systematics
effects in both the temperature and polarization measurements. For
instance, the precise level of power contribution from unresolved
point sources affects $n_{s}$ directly. An independent analysis of
this particular issue by \citet{huffenberger:2006} showed that the
initial point source amplitude quoted by the WMAP team was indeed too
high, which biased $n_{s}$ to low values. Similarly, on large scales
the likelihood approximation used by the WMAP team was biased high
\citep{eriksen:2007}, which also biased $n_s$ low. After these
corrections, the statistical significance of $n_s \ne 1$ dropped to
$\sim2\sigma$.

For polarization the situation may be even more serious due to the
strong sensitivity of $\tau$ and $n_s$ on the low-$\ell$ EE spectrum,
combined with the low signal-to-noise ratio of the WMAP
data. Systematic effects, both from the instrument itself
\citep{jarosik:2007} and from non-cosmological foregrounds
\citep{kogut:2007}, are much more likely to affect the results, and we
are also much less likely to detect them. It is therefore imperative
to carefully check both the data and the analysis methods, in order to
build up confidence in the final cosmological results. In this Letter,
we start this task by computing the low-$\ell$ EE, EB, BB and
foreground template amplitude posterior distributions from the WMAP
data.

\section{Method}
\label{sec:method}

We use a previously introduced Gibbs sampling framework (Jewell et
al.\ 2004, Wandelt et al.\ 2004, Eriksen et al.\ 2004, Larson et al.\
2007; hereafter JWEL) to estimate the posterior distributions. For
full details on the method, we refer the interested reader to the
quoted papers, and only summarize the principles here.

\begin{figure*}[t]

\mbox{\epsfig{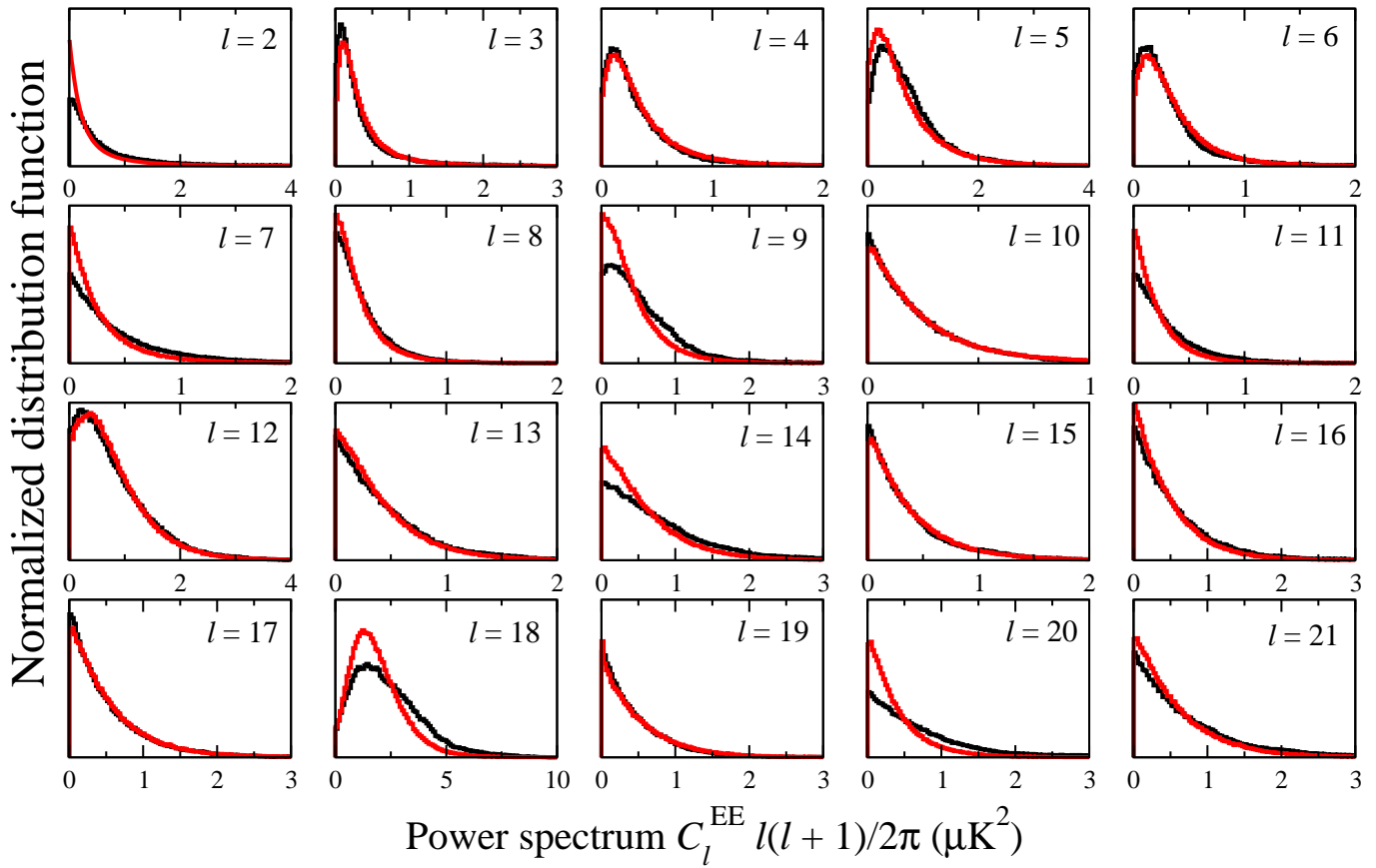}}

\caption{Comparison of single-$\ell$ EE marginal posteriors from the
  WMAP code (black distributions) and the Gibbs sampler (red
  distributions) using the QV map. The EB and BB power spectra are
  held at zero.}
\label{fig:EE_comparison}
\end{figure*}

First we define our notation. The desired distribution is denoted
$P(\mathbf{s}, C_{\ell}, \mathbf{f}|\mathbf{d})$, where $\mathbf{s}$
is the CMB signal, $C_{\ell} = \{C_{\ell}^{\textrm{EE}},
C_{\ell}^{\textrm{EB}}, C_{\ell}^{\textrm{BB}}\}$ is the CMB power
spectrum, $\mathbf{f}$ is a set of foreground template amplitudes, and
$\mathbf{d}$ are the data.

The Gibbs sampler is a Markov Chain Monte Carlo method, and, as such,
maps out the full posterior by drawing samples from it. While direct
evaluation or sampling from the posterior $P(C_{\ell }| \mathbf{d})$
requires inversion of a prohibitively large matrix, the Gibbs sampling
scheme \citep{gelfand:1990} uses the conditional densities of the
joint posterior $P(C_{\ell}, \mathbf{s}|\mathbf{d})$ which is
computationally feasible to sample from. The algorithm may thus be
described by the following sampling steps,
\begin{eqnarray}
  \mathbf{s}^{i+1} & \leftarrow & P(\mathbf{s}|C_{\ell}^{i},
  \mathbf{f}^{i}, \mathbf{d}) \\
  C_{\ell}^{i+1} & \leftarrow & P(C_{\ell}|\mathbf{s}^{i+1},
  \mathbf{f}^{i}, \mathbf{d}) \\
  \mathbf{f}^{i+1} & \leftarrow & P(\mathbf{f}|\mathbf{s}^{i+1},
  C_{\ell}^{i+1}, \mathbf{d}).
\end{eqnarray}
Here the symbol $\leftarrow$ indicates sampling from the conditional
distribution on the right hand side, which can be accomplished without
inverting the signal-plus-noise covariance matrix (see JWEL for
details). For the foreground template amplitude distribution, we note
that the required algorithm is identical to that employed for sampling
monopole and dipole amplitudes \citep{eriksen:2004}.

\section{Data}
\label{sec:data}

We consider only the low-resolution foreground-corrected 3-yr WMAP
polarization data in this Letter, as provided on
LAMBDA\footnote{http://lambda.gsfc.nasa.gov}. These come in the form
of three HEALPix\footnote{http://healpix.jpl.nasa.gov} sky maps,
pixelized at $N_{\textrm{side}} = 16$, each having 3072 pixels in both
Stoke's Q and U. The WMAP P06 sky cut is imposed on the data, leaving
only 2267 pixels for the analysis. Two frequency bands are included,
namely Q-band (41 GHz) and V-band (61 GHz). In addition, we analyze
the co-added map (denoted QV), and also the two frequency
maps jointly but not co-added (denoted Q+V). All maps are provided
with a full noise covariance matrix \citep{jarosik:2007},
appropriately corrected for the P06 sky cut and removal of foreground
templates. The units used in this paper are thermodynamic
$\mu\textrm{K}$.

For foreground marginalization, we adopt two individual
templates. First, we use the K--Ka difference map, smoothed to
$10^{\circ}$ FWHM resolution to reduce noise contributions, as a
tracer of synchrotron emission. Second, for dust emission we adopt the
low-noise template developed by the WMAP team for their foreground
correction procedure \citep{page:2007}. Note that the specific shape
of these templates are of minor importance; if the provided sky maps
are free of foregrounds, they should not correlate significantly with
any non-CMB map.

We compare our results to the official WMAP likelihood
code\footnote{version v2p2p2}, also available from LAMBDA. To map out
the appropriate posteriors, we have written a special-purpose MCMC
wrapper around this likelihood code.

\section{Results}
\label{sec:results}

\subsection{Numerical verification of the WMAP likelihood}
\label{sec:verification}

The first case considered is that adopted by the WMAP likelihood code,
namely the co-added QV map. For this analysis, we fix the EB and BB
spectra to zero, and map out the corresponding marginalized EE
posteriors $\ell$-by-$\ell$, both with the Gibbs sampler and by the
WMAP-based MCMC code.

\begin{figure}[tb]

\mbox{\epsfig{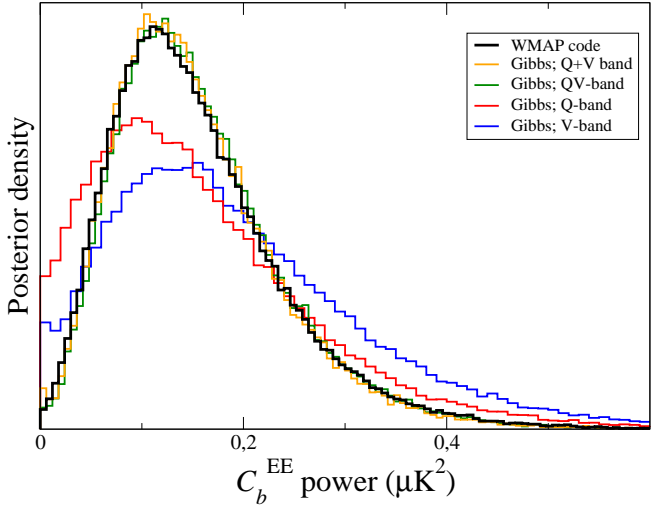}}

\caption{Marginal posterior distributions for the EE bin power between
$\ell=2$--6.} 
\label{fig:EE_bin_power}
\end{figure}

The results from this exercise are shown in Figure
\ref{fig:EE_comparison}. The agreement between the two approaches is
very good, and this is an important validation of the WMAP data
processing method: First, we analyze the data at their native
$N_{\textrm{side}} = 16$ resolution, while the WMAP team analyze maps
downgraded to $N_{\textrm{side}} = 8$. Second, they marginalize over a
single total foreground template, while we marginalize over the K--Ka
difference map and a dust template. Third, we use a Gibbs sampler for
the numerical work, while the WMAP team uses a brute-force likelihood
evaluator. None of these differences affects the low-$\ell$ EE
spectrum peak visibly, as will be quantified more precisely in the
next section.

\subsection{Generalized analysis}
\label{sec:stability}

\begin{deluxetable}{lcccc}
\tablewidth{0pt} 
\tabletypesize{\small} 
\tablecaption{Marginalized EE and BB band powers\label{tab:band_power}}
\tablecolumns{5}
\tablehead{ & \multicolumn{2}{c}{EE power ($10^{-1}\,\mu\textrm{K}^2$)} &
  \multicolumn{2}{c}{BB power ($10^{-1}\,\mu\textrm{K}^2$)} \\ Data set & $\ell=2$--6 & $\ell=2$--20 &$\ell=2$--6 & $\ell=2$--20 
}

\startdata
\cutinhead{EE free; EB = BB = 0}
WMAP & $1.1^{+0.9}_{-0.5}$ & $0.64^{+0.46}_{-0.34}$ & \nodata & \nodata \\
QV-band & $1.2^{+0.9}_{-0.6}$ & $0.67^{+0.39}_{-0.38}$ &
\nodata & \nodata \\
Q+V-band & $1.1^{+0.8}_{-0.6}$ & $0.65^{+0.38}_{-0.35}$ &
\nodata & \nodata \\
Q-band & $1.0^{+1.0}_{-0.8}$ & $0.36^{+0.67}_{-0.36}$ &
\nodata & \nodata \\
V-band & $1.3^{+1.2}_{-0.9}$ & $1.2^{+0.9}_{-0.7}$ &
\nodata & \nodata \\

\cutinhead{EE, BB free; EB = 0}
WMAP & $0.94^{+0.76}_{-0.58}$ & $0.63^{+0.44}_{-0.37}$ &
$<0.70$ & $<0.40$  \\
QV-band & $1.1^{+0.8}_{-0.6}$ & $0.61^{+0.38}_{-0.37}$ &
$<0.57$ & $<0.26$ \\
Q+V-band & $1.1^{+0.8}_{-0.6}$ & $0.57^{+0.40}_{-0.31}$ &
$<0.58$ & $<0.30$ \\
Q-band & $0.3^{+1.3}_{-0.3}$ & $0.23^{+0.68}_{-0.23}$ &
$0.3^{+1.2}_{-0.3}$ & $<0.71$ \\
V-band & $1.4^{+1.4}_{-0.9}$ & $1.1^{+1.0}_{-0.7}$ &
$<0.94$ & $<0.51$ \\

\cutinhead{EE, EB, BB free}
QV-band & $1.4^{+0.9}_{-0.7}$ & $0.65^{+0.41}_{-0.30}$ &
$0.30^{+0.60}_{-0.30}$ & $0.1^{+0.3}_{-0.1}$ \\
Q+V-band & $1.3^{+0.9}_{-0.7}$ & $0.66^{+0.43}_{-0.30}$ &
$0.31^{+0.59}_{-0.31}$ & $0.1^{+0.3}_{-0.1}$ \\
Q-band & $1.1^{+1.1}_{-0.8}$ & $0.54^{+0.65}_{-0.36}$ &
$0.7^{+1.3}_{-0.6}$ & $0.5^{+0.7}_{-0.4}$ \\
V-band & $1.8^{+1.6}_{-1.1}$ & $1.5^{+0.9}_{-0.9}$ &
$0.47^{+0.93}_{-0.47}$ & $0.4^{+0.6}_{-0.4}$ 

\enddata

\tablecomments{Values indicate either the posterior mode and upper
and lower 68\% confidence interval, or the upper 68\% confidence
limits. If the lower error bar equals the posterior mode value, a
non-zero peak is detected, but at a significance of less than
68\%.}

\end{deluxetable}

Having validated the numerics and data processing steps in both
procedures, we now expand the analysis and allow for non-zero EB
and/or BB spectra. We also compute the posteriors for each frequency
band separately and jointly. For presentational reasons, we report
only band powers in EE and BB between $\ell=2$--6 and $\ell=2$--20,
respectively. Cases of special interest are treated separately in
subsequent sections. 

In order to achieve good convergence, $10^6$ samples were generated
for the non-zero EB, $\ell=2$--20 cases. For all other cases, $10^5$
samples were generated. The CPU time to generate one sample was $\sim2$
seconds.

The results from these computations are summarized in Table
\ref{tab:band_power}. The EE posteriors with fixed EB = BB = 0 are
shown in Figure \ref{fig:EE_bin_power}. Again, note the excellent
agreement between the WMAP results and the QV and Q+V cases in the two
top sections.

However, even though the joint QV analyses agree well, the picture is
considerably less clear when it comes to single bands and relaxed EB
constraints. On the one hand, there appears to be more EE power in the
V-band data than in the Q-band data. On the other hand, there may seem
to be small traces of BB power in the Q-band data.

The BB posteriors develop a peak away from $C^{\textrm{BB}}_{\ell}=0$
in the EB$\ne0$ case. This is not surprising. Since the signal
covariance matrix is positive definite, one must have
$C_{\ell}^{\textrm{EE}} C_{\ell}^{\textrm{BB}} >
(C_{\ell}^{\textrm{EB}})^2$. Therefore, when marginalizing over
$C_{\ell}^{\textrm{EB}}$, a non-zero peak emerges in both the EE and
BB spectra individually.

\subsubsection{Foreground amplitude posteriors}

\begin{figure*}[t]

\mbox{\epsfig{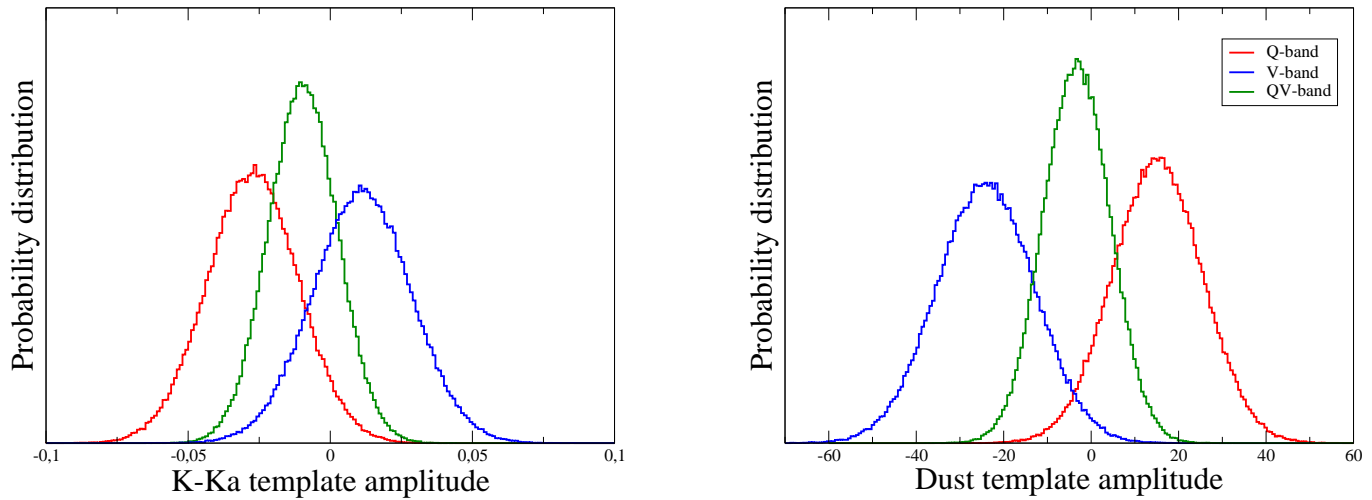}}

\caption{The foreground template amplitude marginal posteriors for the
  Q-, V- and QV-bands. For this plot, the EB and BB power spectra were
  set to zero.}
\label{fig:foregrounds}
\end{figure*}




In Figure \ref{fig:foregrounds} we show the foreground template
amplitude posteriors for the Q-, V- and QV-band data, for the case
with fixed EB = BB = 0. Although no signs of significant residual
foregrounds are observed in the co-added QV band, apparently
confirming the fits made by the WMAP team, the same is not true for
each band individually. On the contrary, non-zero correlations are
seen in both the Q- and V-band data individually. 

For the Q-band data, the marginal best-fit K--Ka amplitude is
$A_{\textrm{s}} = -0.027^{+0.014}_{-0.17}$, different from zero at
$2\sigma$. The best-fit dust amplitude is $A_{\textrm{d}} =
15.7^{+8.7}_{-10.7}$. For the V-band data, the best-fit dust amplitude
is $A_{\textrm{d}} = -24.1^{+10.3}_{-11.3}$, 2.3$\sigma$ away from
zero. The K--Ka amplitude is $A_{\textrm{s}} =
0.011^{+0.015}_{-0.018}$.

These results may be compared to Table 4 of \citet{page:2007}. The
main difference between the two analyses is that while we apply the
conservative P06 mask to the data, \citet{page:2007} apply the much
more aggressive processing mask described by \citet{jarosik:2007}. The
two analyses remove 26.4\% and 5.7\% of the sky,
respectively. Considering that all cosmological analyses are carried
out with the P06 mask, and that variations in the synchrotron spectral
index are observed between the galactic plane and the high latitudes
\citep{kogut:2007}, it is not immediately clear to us why the more
aggressive mask was chosen for this task by the WMAP team. The
improvement in raw $\chi^2$ after further correcting the ``cleaned''
WMAP maps for these residuals is -5.4 for the Q-band and -3.3 for the
V-band, respectively.

Two other, minor differences are that we take into account both the
full noise and CMB covariance matrices, while the WMAP team includes
only a white noise covariance approximation. Note that this carries no
extra cost within the Gibbs sampling framework.

\subsubsection{ E$\times$B cross-correlation spectra}

\begin{figure}[t]

\mbox{\epsfig{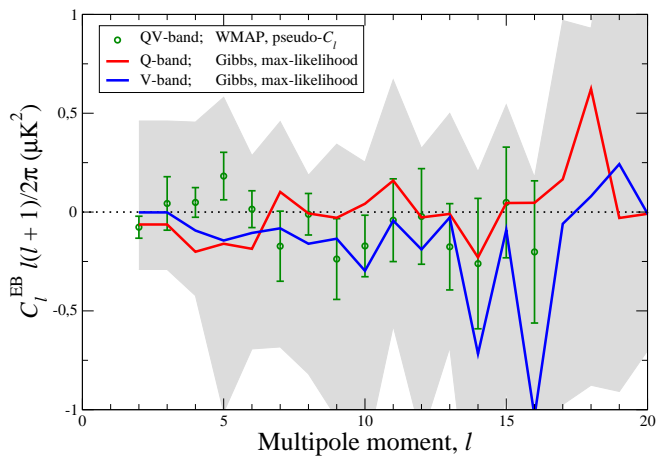}}

\caption{The E$\times$B cross-power spectrum. The gray region shows
  the $1\sigma$ confidence region around the V-band spectrum (blue
  curve); note that the marginal EB posterior is strongly non-Gaussian
  with heavy tails. The green data points show the WMAP EB spectrum
  computed with a pseudo-$C_{\ell}$ estimator with Gaussian error
  bars.}
\label{fig:EB_correlation}
\end{figure}


We find evidence for a non-zero EB correlation in the V-band sky map.
This spectrum is shown in Figure
\ref{fig:EB_correlation} for the Q- and V-band sky maps individually,
together with the pseudo-$C_{\ell}$ EB QV-band spectrum computed by
the WMAP team. Note the consistently negative correlation observed in
the V-band. Although the significance of the negative correlation is
not more than a few tenths per multipole, and the joint significance
is not more than 1 to 2$\sigma$ depending on binning scheme, it is
observed consistently in every multipole up to $\ell=17$. A similar,
though weaker, trend is seen in the pseudo-spectrum computed by the
WMAP team from the QV combination.

\section{Conclusions}
\label{sec:conclusions}

We have performed a Bayesian analysis of the low-resolution
foreground-corrected 3-yr WMAP polarization data using a previously
described methodology based on the Gibbs sampler. By doing so, we
validated the numerical implementation of the official WMAP likelihood
code, as well as their procedure for degradation of map
resolution. However, when generalizing the analysis to allow for
non-zero EB and BB power spectrum components, and also considering
individual frequency bands, we found several issues that may be
important for the cosmological interpretation of these data.

First and foremost, when relaxing the constraints on EB and BB,
noticeable differences between the Q- and V-band posteriors are
observed. Specifically, we find generally more EE power in the V-band
data than in the Q-band data, but also perhaps some hints of BB power
in the Q-band data. At the same time, we have also found a negative EB
correlation in the V-band map, as well as residual foregrounds in both
maps.

If these tentative findings are confirmed by future experiments or
additional years of WMAP observations, significant shifts in
cosmological parameters could be the result. For example, if the
V-band data alone were used for the WMAP3 analysis instead of the QV
combination, the amplitude of the $\ell=2$--6 EE detection would increase
by 20--50\%, depending on whether BB is allowed to vary or
not. Consequently, $\tau$ could increase from 0.09 to $\sim0.12$, and
$n_{\textrm{s}}$ by a percent or two, comparable to its current
nominal statistical uncertainty of 0.015, from $\sim0.97$ to
$\sim0.98$.

As discussed in the introduction, systematics are a serious concern
for both the temperature and polarization data for both $\tau$ and
$n_s$. It is important to bear in mind that the currently quoted
uncertainties on these quantities often found in the literature are
statistical only. The unknown systematic uncertainties may turn out to
be non-negligible for the currently available data sets, and, in
particular, we believe it is too early to draw any firm conclusions
concerning the precise value of $n_s$. Fortunately, Planck will
clarify these issues in the near future.

\begin{acknowledgements}
  We wish to thank David Larson for his contributions during the early
  phases of this project. We acknowledge use of the Legacy Archive for
  Microwave Background Data Analysis (LAMBDA). We acknowledge use of
  the HEALPix software \citep{gorski:2005} and analysis package for
  deriving the results in this paper. This work was partially
  performed at the Jet Propulsion Laboratory, California Institute of
  Technology, under a contract with the National Aeronautics and Space
  Administration. HKE acknowledges financial support from the Research
  Council of Norway. BDW acknowledges support by NSF grant
  no. 0507676, the Friedrich Wilhelm Bessel Research Prize by the
  Alexander v. Humboldt foundation and NASA subcontract no. 1236748.
\end{acknowledgements}






\end{document}